\documentclass[acmlarge, nonacm, screen]{acmart}

\acmConference[GenAICHI]{The The CHI 2022 Workshop on Generative AI and HCI} % need to include this to suppress addresses footer

\newcommand\extrafootertext[1]{%
    \bgroup
    \renewcommand\thefootnote{\fnsymbol{footnote}}%
    \renewcommand\thempfootnote{\fnsymbol{mpfootnote}}%
    \footnotetext[0]{#1}%
    \egroup
}
\newcommand{\workshopname}{GenAICHI: CHI 2022 Workshop on Generative AI and HCI}
\newcommand{\licensedetails}{Licensed under a Creative Commons Attribution 4.0 International License (CC BY 4.0). Copyright remains with the author(s).}

\AtBeginDocument{
    \fancypagestyle{firstpagestyle}{
        \fancyhf{}
        \fancyfoot[L]{\sffamily\footnotesize \workshopname}%
        \fancyfoot[C]{\sffamily\footnotesize \thepage}
    }
    \fancyhf{}
    \fancyhead[L]{\sffamily\footnotesize\shorttitle}
    \fancyhead[R]{\sffamily\footnotesize\shortauthors}
    \fancyfoot[L]{\sffamily\footnotesize\workshopname}%
    \fancyfoot[C]{\sffamily\footnotesize\thepage}
    \extrafootertext{\licensedetails}
}

\usepackage{caption}
\usepackage{subcaption}
\AtBeginDocument{%
  \providecommand\BibTeX{{%
    \normalfont B\kern-0.5em{\scshape i\kern-0.25em b}\kern-0.8em\TeX}}}

\begin{document}

\title{Embodying the Glitch: Perspectives on Generative AI in Dance Practice}

\author{Benedikte Wallace}
\affiliation{%
  \institution{University of Oslo}
  \city{Oslo}
  \country{Norway}}
\email{benediwa@ifi.uio.no}

\author{Charles Patrick Martin}
\affiliation{%
  \institution{Australian National University}
  \city{Canberra}
  \country{Australia}}

\renewcommand{\shortauthors}{Wallace and Martin}

\begin{abstract}
  What role does the break from realism play in the potential for generative artificial intelligence as a creative tool?
  Through exploration of \textit{glitch}, we examine the prospective value of these artefacts in creative practice.
  This paper describes findings from an exploration of AI-generated "mistakes" when using movement produced by a generative deep learning model as an inspiration source in dance composition.
\end{abstract}


\begin{teaserfigure}
\center
  \includegraphics[scale=0.12]{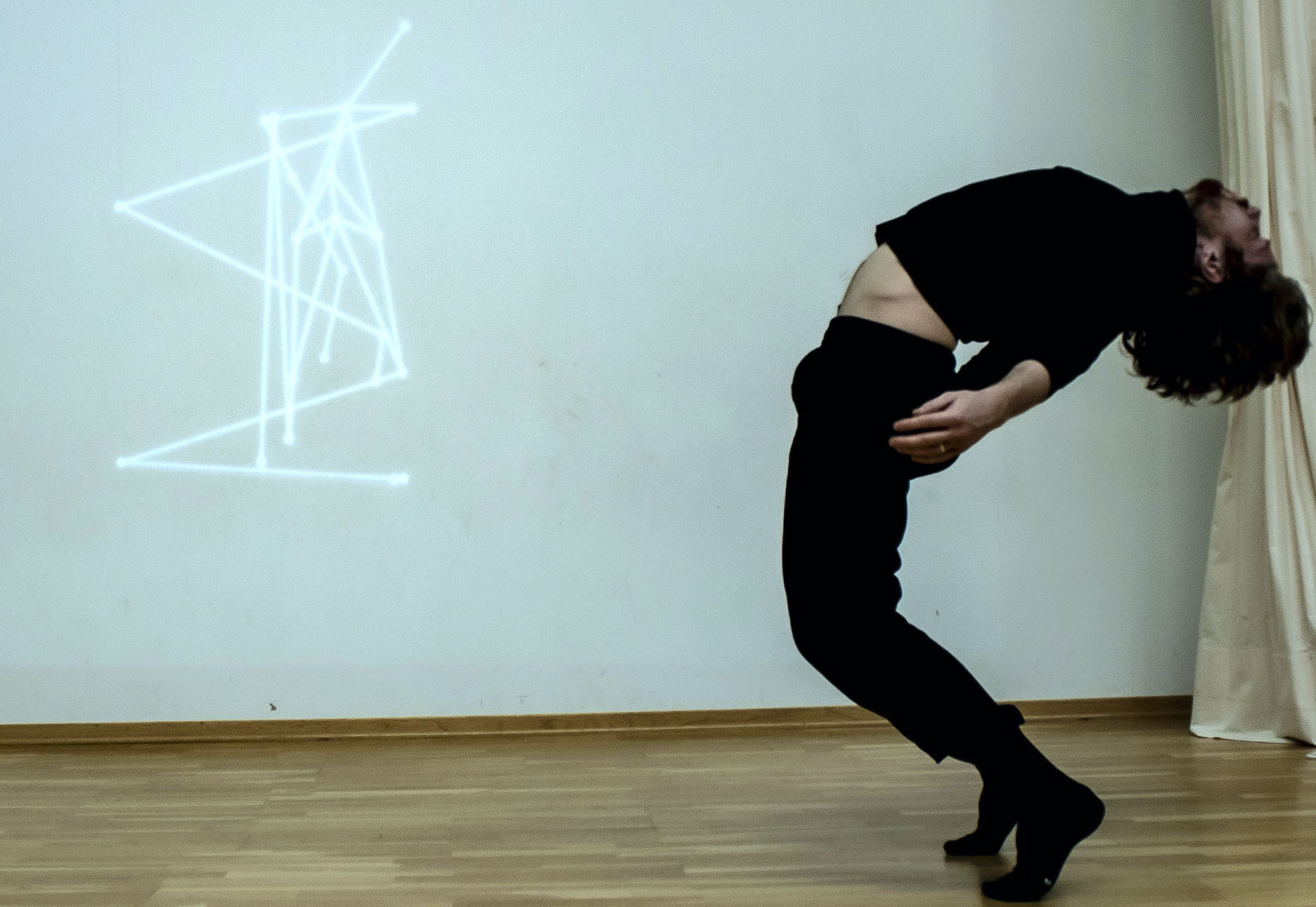}
%   \caption{Dancer and choreographer Bendik Sundby exploring the glitch}
\caption{A dancer and choreographer exploring the glitch}
  \label{fig:teaser}
\end{teaserfigure}

\maketitle

\section{Movement Inspiration from AI}
%Why did the dancers choose to come and dance for my dataset? how did they expect a thing like this to be useful to them? XXXX

In our work with generative AI models of human movement, we have come across many examples of models failing to produce movements that are human-like. 
We refer to these "failings" as \textit{glitches}. 
These glitches can take many different forms. 
Sometimes the models get stuck in a loop of movement that it keeps repeating, other times the output becomes completely unrecognizable as a human form.
Occasionally the output falls in between noise and repetition, %morphing between a human-like shape and noise, 
slipping into chaos and re-emerging from it. 
In previous work, we trained several generative deep neural networks on a data set of 3D motion capture data of improvised dance.
Our implementation does not include any inbuilt movement constraints, instead, the models need to learn these constraints from the underlying training data. 
As there are also no corrections applied to the generated output, the models will occasionally produce impossible movements. 
Limbs can extend and rotate past human limits at velocities that would break physical laws. 
We often include examples of these glitches in presentations, both to show how the model improves during training and how different sampling strategies affect the types of glitches that are produced \cite{wallaceExploring}.
Interestingly, the audiences that are most drawn to these "mistakes" are those who are themselves dancers and movers. 
Something about these impossible, yet somehow recognizable shapes seem to prompt movement in these audiences. It is as though they are trying to make sense of the glitches through their bodies.
It might initially make sense to assume that individuals who use movement in their artistic practice and are very much aware of their own bio-mechanical limits would be the model's harshest critics. 
Instead, what we found was perhaps closer to the opposite. 

As our work aims to produce models that might contribute to the creative processes of dancers, we seek to understand the relationship between artists and generative AI.
The reoccurring positive response from dancers to the model's apparent failings gave rise to the idea of glitch tolerance in movement generation when the rationale for the generated output is to spark artistic inspiration and novel ideas. 
We pose the question: to what extent could the glitches inspire movement and contribute to the dancer's creative task?
In this work, we attempt to examine this notion through a movement experiment.
We presented a collection of movement sequences generated by deep neural networks to a dancer for a dance composition task. The movement clips range from clearly humanoid to completely abstract.
Through our discussions with the dancer following the experiment we examine what level of glitch is most rewarding and why this break from realism might contribute to his improvisation in a positive way.

\section{Glitch tolerance}
Naturally, deviance from realism is not desirable in all cases.
When generating realistic movement sequences for animation, for example, generative methods with high realism have already shown great promise in decreasing the workload of animators \cite{harvey2020robust,starke2020local}.
The realism and accuracy of a generative AI can also be important in interactions where it is beneficial for the user to gain a feeling of control \cite{fdili2019making}.
%
%
%But perhaps, from a creative process or performance  perspective, it is more interesting when weird things happen - like googles weird walker \cite{merel2017learning}
%
%Perhaps it has its place in applications that should function as suggestion givers. 
%
Even within the domain of creativity support, generative AI mistakes seem to hold different levels of latent quality.
Practitioners and audiences seem to have varying tolerance for glitches depending on art form and genre.
Why then do dancers and performance artists seem to be among the glitch tolerant?

In contrast to music and writing, dance involves strict biological constraints. %
We are restricted by what the human body can do. 
While composers and writers can randomise characters or notes as easily as a generative AI, dancers cannot grow or shrink their limbs, and their speed and position are limited.
Perhaps it is exactly these inbuilt limitations that make dance more glitch tolerant. 
In order to transform the glitches produced by the AI into something the body can do, an expansion of our interpretation is required as the dancer attempts to embody the glitch.
%. 
%In this way the glitches may allow for ideas that are on the boundaries of what is possible to come through. 
%
In this way, the glitches are similar to how dancers and performance artists might use inanimate objects in performances or in their creative process \cite{galeanoObjects}. 
The task becomes transmodal in the attempt to imbue objects with meaning and movement.
This transmodality can be likened to tasks such as converting text to images\footnote{\href{https://twitter.com/RiversHaveWings}{https://twitter.com/RiversHaveWings}}, or creating drum patterns using a Chopin {\'e}tude\footnote{\href{https://youtu.be/wY1dIY--UDY?t=366}{https://youtu.be/wY1dIY--UDY?t=366}}. 
%
%Dance involves multifactorial sensing of your environment, the space, other bodies or objects, sounds, and your own biological affordances and limitations. 
%
%A complex interweaving of past and present impressions affect the movements a dance artist conveys during improvisation and choreographic exploration. 
%
%Through their movements dancers both disseminate and are affected by notions of time, space and emotion. 
%
%Taking inspiration from inanimate objects, music, words etc.
%
%Quote from Diego may be relevant to my point here?: Sometimes we make choreos with objects. 
%
%By putting something into the object to make it be alive. it develops its own personality, you develop feelings towards it. While the objects arent usually giving input to you in the form of movement, but it has a shape, a feeling?  
%
%For example working with masks. Or puppets. How do you imbue “dead” things with life? 

%
\section{Embodying the glitch}
\begin{figure}
     \centering
     \begin{subfigure}[b]{.45\linewidth}
         \centering
         \includegraphics[width=\linewidth]{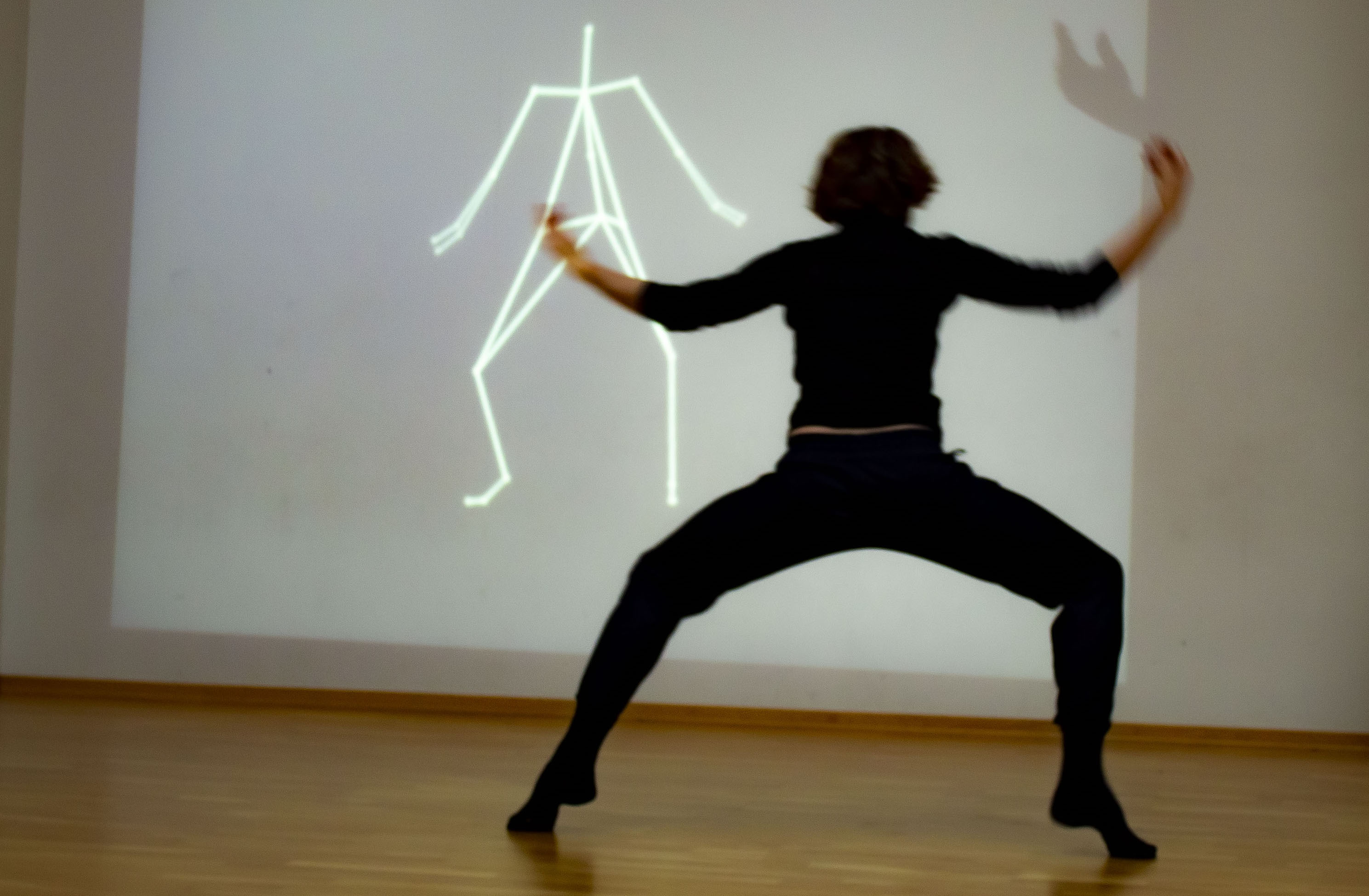}
         \caption{}
         \label{fig:norm}
     \end{subfigure}
     %\hfill
     \begin{subfigure}[b]{.45\linewidth}
         \centering
         \includegraphics[width=\linewidth]{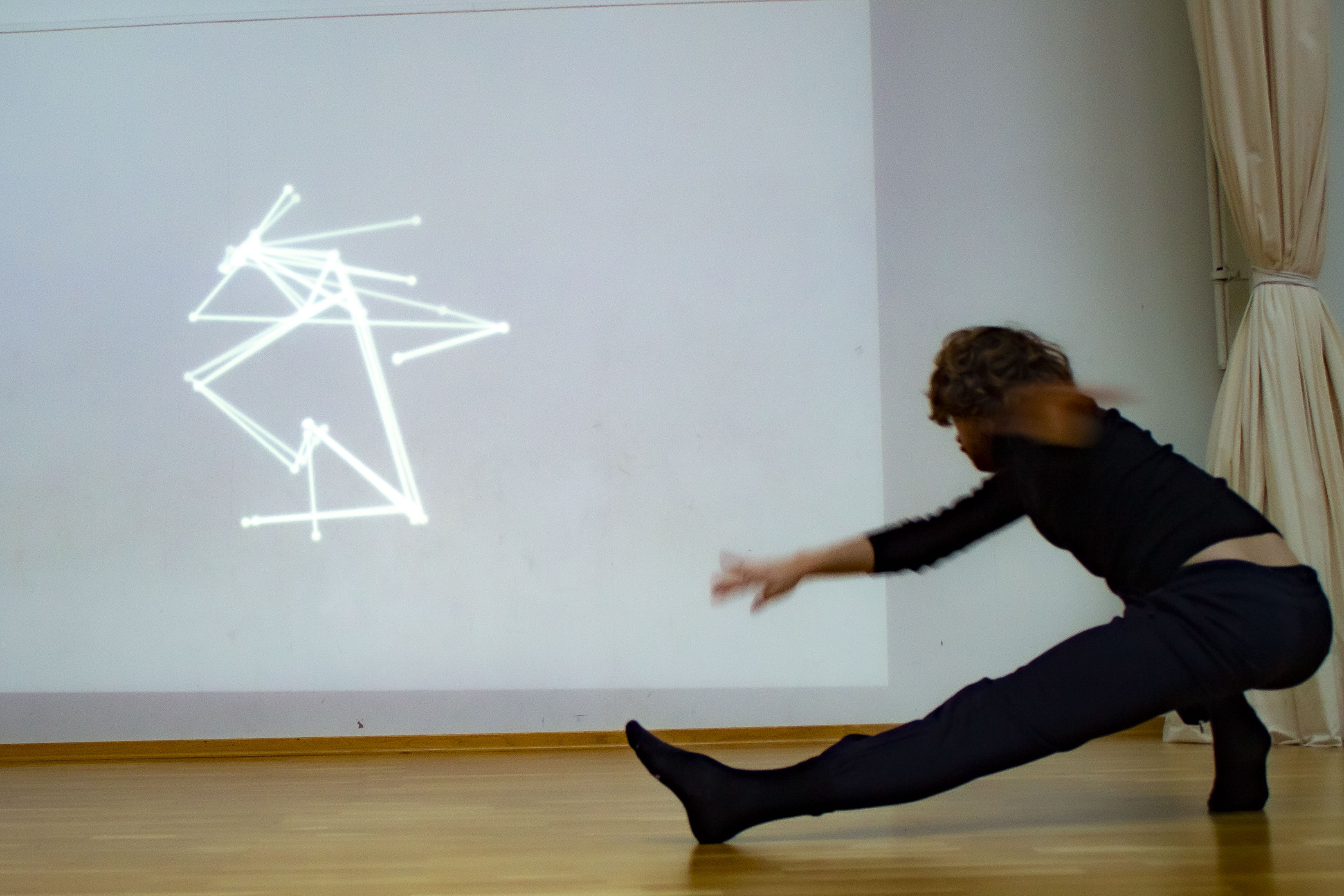}
         \caption{}
         \label{fig:glitch}
     \end{subfigure}
        \caption{The dancer watches and reacts to the movement sequences generated by the model. Some sequences show a clear humanoid shape %\ref{fig:norm}
        while others are more abstract. %\ref{fig:glitch}
        }
        \label{fig:twobendik}
\end{figure}

\begin{figure}
    \centering
    \begin{subfigure}[b]{0.45\linewidth}
        \includegraphics[scale=.18]{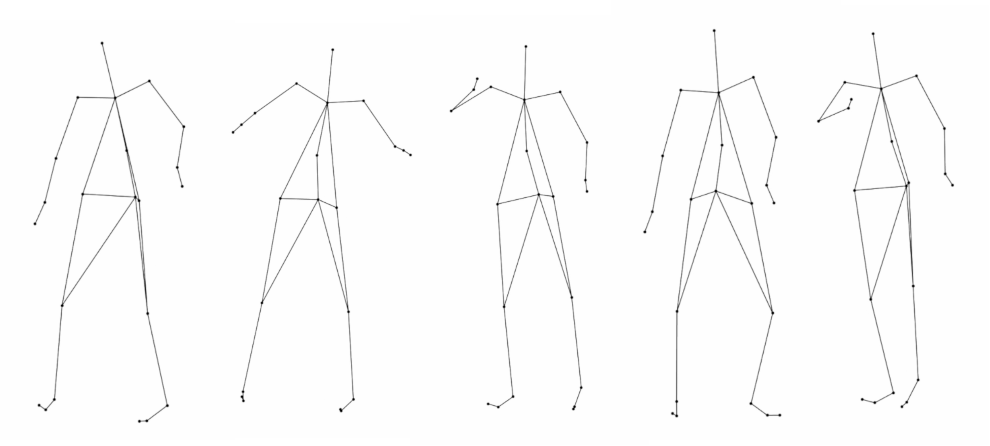}
        \caption{}
        \label{fig:a}
    \end{subfigure}
    \vfill
    \begin{subfigure}[b]{0.45\linewidth}
        \includegraphics[width=1.1\linewidth]{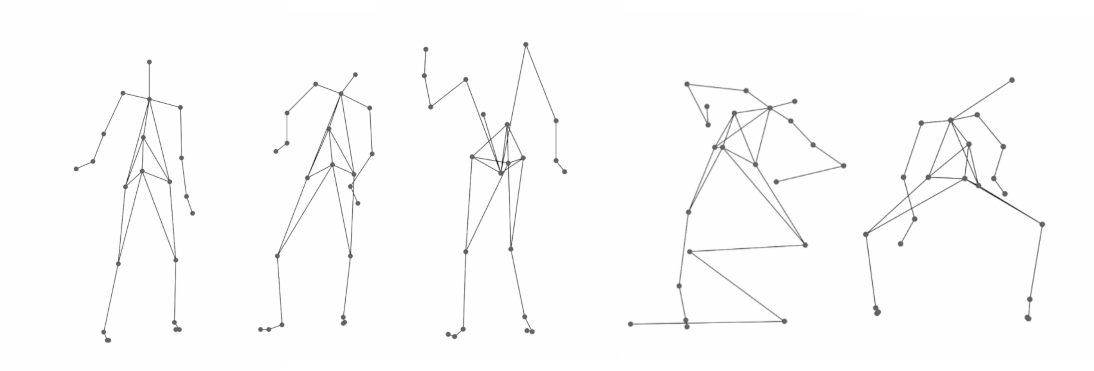}
        \caption{}
        \label{fig:b}
    \end{subfigure}
    \hfill
    \begin{subfigure}[b]{0.45\linewidth}
        \includegraphics[width=1.1\linewidth]{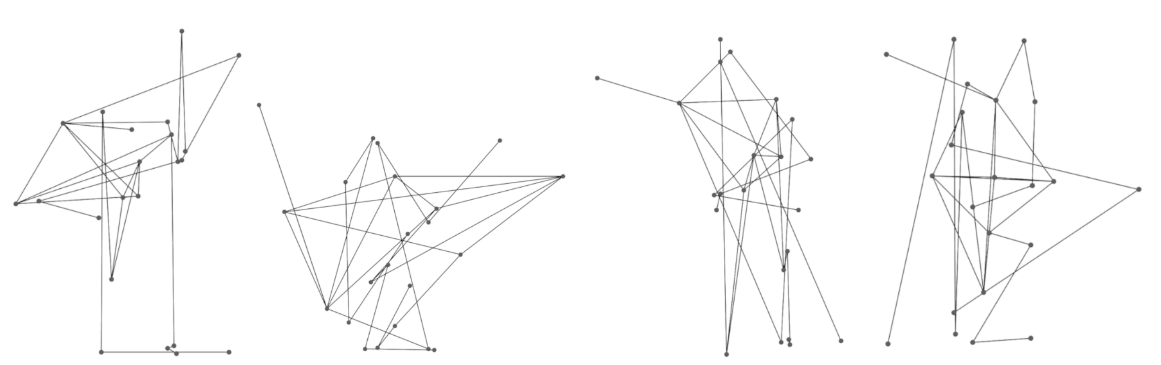}
        \caption{}
        \label{fig:c}
    \end{subfigure}
    \caption{
        The sequences range from stable, human-like movements (\subref{fig:a}) to semi-glitched, where the sequence contains moments where the body distorts (\subref{fig:b}) to fully glitched (\subref{fig:c}) where it is no longer possible to differentiate between limbs. 
    }
    \label{fig:glitch-examples}
\end{figure}

We conducted an experiment wherein a dancer was invited to browse a series of AI-generated movement sequences, dance in response and discuss their experience. 
The dancer was asked to imagine that he needs to compose a dance and that his inspiration for the piece can come from 1 or more of 10 short (10-30 seconds) clips.
The majority of the sequences were generated by a Mixture Density Recurrent Neural Network (MDRNN) trained on a dataset of improvised dance.
For additional details on the data set and MDRNN implementation, we refer readers to \cite{wallaceLearning}.
The remaining examples were produced by a Transformer model \cite{vaswani2017attention} trained on the same data. 
The movement sequences were shown using point-line visualisations projected onto the wall of the dance studio (See Figure \ref{fig:twobendik}). 
Each of the 10 clips was played until the dancer wished to move on to the next clip. 
%
%The clips range from abstract shapes where the human form is not identifiable to simple movement sequences such as swaying and small arm movements. 
%
%Some clips show a mix between the two where the model has produced sequences that are human-like in some parts and distorted in others. 
%
%In these clips limbs twist, bend, stretch or shrink in unrealistic ways. 
%
Figure \ref{fig:glitch-examples} shows examples of 3 different clips used in this session.

\section{Reflections on movement}
The movement session described above was followed up with a discussion where the dancer and first author reflect on the experience. 
A strategy that became apparent during the movement session is that the dancer at times seems to mirror or shadow \cite{blackwell2012live} the movements displayed by the AI. 
This mirroring may be the most recognizable simply because it is the clearest transformation from what the dancer sees and what he subsequently portrays with his own body. 
This phenomenon is naturally only clear when the dancer is interacting with the examples that are the most realistic.
While the dancer mentioned being automatically drawn to the movement patterns of the extremities of the shape (arms and legs), he specified that the entire form, its shape, dynamic flow and movement direction affects his improvisation. 
He explained that the clips where the human shape is clearly visible lay stricter guidelines for where his movements begin and where they go.
In this sense, the more humanoid shapes portraying simple movements such as extending an arm are easier to work with as they form a good foundation for his improvisation.
However, when imagining using a generative model like this to browse through movement sequences and see if something sparks his imagination, he clarified that he would not necessarily want all sequences to be recognizable as a human form.
While the movements that are clearly discernible as human are easy for the dancer to incorporate or translate into his own movement patterns, the more abstract shapes trigger other thoughts.
He is then drawn to the overall shape of the movement to a higher degree. How does it rotate? Are the movements soft or sharp? Does it come across as stretched out or compressed?
The dancer expresses that these glitches in the shape allow him to feel like he is being more innovative and "personal" in his own movements compared to when the AI is less abstract.
The clips the dancer interacted with the longest and enjoyed the most were those that morph back and forth between a clearly discernible human form and more abstract shapes or impossible movements. 
The human-like movements ground his movements in the concrete. 
When the figure distorts, shrinks and grows, he can no longer mimic the shape of the body he sees and thus his focus shifts to other aspects and allows for new ideas to form.

\section{Findings}
%to what extent could the glitches inspire movement and contribute to the dancer's creative task?
Our study suggests that the AI-generated movements were able to draw the dancer out of his usual movement patterns in different ways.
The dancer explained that the glitches made him feel more innovative and they also allowed him to be more personal, reverting perhaps more into himself and away from interaction with the AI. 
The more human-like examples gave the dancer grounding in concrete movement concepts. 
He describes these clips as providing a clear starting point for improvisation. 
Concurrently, they introduce a mental constraint that can lead to disinterest. %
The glitches appear to be most beneficial when they are transient. 
In these cases, the dancer finds both a grounding and an inspirational goal that is unreachable.
This invites him to attempt to translate abstract shapes or impossible movements into those that he can perform. 
This research describes an experiment to examine how a single dancer with a background in lyrical modern dance responds to glitches in our generative AI system.
Other dancers with different backgrounds could experience the experiment differently. 
In future work, we intend to further examine the expectations and hopes dancers and choreographers have for generative AI in their practice by facilitating a workshop involving dancers from various movement backgrounds.
%
%From this study we conclude
Our work indicates that realism may not be the ultimate goal for generative AI in an artistic practice. Instead, we suggest that there is potential in occasionally embracing, and even embodying, the glitch.

\begin{acks}
The authors would like to thank dancer and choreographer Bendik Sundby for contributing his time, reflections and performance.
This work was partially supported by the Research Council of Norway through its Centres of Excellence scheme, project number 262762.
\end{acks}

\bibliographystyle{ACM-Reference-Format}
\bibliography{bib}
\end{document}